\documentclass[11pt]{article}
\usepackage{amssymb,amsmath,fullpage,graphicx}
\usepackage{enumerate}
\def\x{\mathbf x}

\makeatletter \numberwithin{equation}{section}

\begin{document}
\title{Stability switching at transcritical bifurcations of solitary waves in generalized nonlinear Schr\"odinger equations}
\author{Jianke Yang \\
Department of Mathematics and Statistics \\ University of Vermont\\
Burlington, VT 05401, USA}
\date{ }
\maketitle

\begin{abstract}
Linear stability of solitary waves near transcritical bifurcations
is analyzed for the generalized nonlinear Schr\"odinger equations
with arbitrary forms of nonlinearity and external potentials in
arbitrary spatial dimensions. Bifurcation of linear-stability
eigenvalues associated with this transcritical bifurcation is
analytically calculated. Based on this eigenvalue bifurcation, it is
shown that both solution branches undergo stability switching at the
transcritical bifurcation point. In addition, the two solution
branches have opposite linear stability. These stability properties
closely resemble those for transcritical bifurcations in
finite-dimensional dynamical systems. This resemblance for
transcritical bifurcations contrasts those for saddle-node and
pitchfork bifurcations, where stability properties in the
generalized nonlinear Schr\"odinger equations differ significantly
from those in finite-dimensional dynamical systems. The analytical
results are also compared with numerical results, and good agreement
is obtained.
\end{abstract}




\section{Introduction}
The generalized nonlinear Schr\"odinger equations considered in this
article are a large class of Schr\"odinger-type equations which
contain arbitrary forms of nonlinearity and external potentials in
arbitrary spatial dimensions. This class of equations are physically
important since they include theoretical models for nonlinear light
propagation in refractive-index-modulated optical media
\cite{Kivshar_book,Yang_SIAM} and for atomic interactions in
Bose-Einstein condensates loaded in magnetic or optical traps
\cite{BEC} as special cases. Given their physical importance, these
equations have been heavily studied in the physical and mathematical
communities. These equations admit a special but important class of
solutions called solitary waves, which are spatially localized and
temporally stationary structures of the system. These solitary waves
exist for continuous ranges of the propagation constant. At special
values of the propagation constant and under certain conditions,
bifurcations of solitary waves can occur. Indeed, various solitary
wave bifurcations in these equations have been reported. Examples
include saddle-node bifurcations (also called fold bifurcations)
\cite{Yang_SIAM,Panos_2005,Kapitula_2006,Panos_PhysicaD_2009,Akylas_2012,Yang_saddle1,Yang_saddle2},
pitchfork bifurcations (sometimes called symmetry-breaking
bifurcations)
\cite{Akylas_2012,Malomed_pitchfork_2007,Weinstein_2008,Sacchetti_2009,Panos_2009,Kirr_2011},
and transcritical bifurcations \cite{Yang_classification}. These
three types of bifurcations have also been classified in
\cite{Yang_classification}, where analytical conditions for their
occurrence were derived.

Stability of solitary waves near these bifurcations is an important
issue. In finite-dimensional dynamical systems, stability of
fixed-point branches near these bifurcations is well known
\cite{GH}. However, as was explained in
\cite{Yang_saddle2,Yang_pitchfork}, those stability results from
finite-dimensional dynamical systems cannot be extrapolated to the
generalized nonlinear Schr\"odinger equations. Thus this stability
in the generalized nonlinear Schr\"odinger equations has to be
studied separately. For saddle-node bifurcations of solitary waves,
this stability question has been analyzed in
\cite{Yang_saddle1,Yang_saddle2}. It was shown that no stability
switching takes place at a saddle-node bifurcation, which is in
stark contrast with finite-dimensional dynamical systems where
stability switching generally takes place \cite{GH}. For pitchfork
bifurcations  of solitary waves, this stability has been analyzed in
\cite{Weinstein_2008,Sacchetti_2009,Kirr_2011,Yang_pitchfork}. It
was shown that this stability possesses novel features which have no
counterparts in finite-dimensional dynamical systems as well. For
instance, the base and bifurcated branches of solitary waves (on the
same side of a pitchfork bifurcation point) can be both stable or
both unstable \cite{Kirr_2011,Yang_pitchfork}, which contrasts
finite-dimensional dynamical systems where the bifurcated
fixed-point branches generally have the opposite stability of the
base branch \cite{GH}. For transcritical bifurcations of solitary
waves, the stability question is still open.

In this paper, we analyze the linear stability of solitary waves
near transcritical bifurcations in the generalized nonlinear
Schr\"odinger equations. Our strategy is to analytically calculate
bifurcations of linear-stability eigenvalues from the origin when
the transcritical bifurcation occurs. Based on this eigenvalue
bifurcation and assuming no other instabilities interfere, linear
stability of solitary waves near the transcritical bifurcation point
is then obtained. We show that both solution branches undergo
stability switching at the transcritical bifurcation point. In
addition, the two solution branches have opposite linear stability.
These stability properties closely resemble those for transcritical
bifurcations in finite-dimensional dynamical systems. Thus, among
the three major bifurcations (i.e., saddle-node, pitchfork and
transcritical bifurcations), the transcritical bifurcation is the
only one where stability properties in the generalized nonlinear
Schr\"odinger equations closely resemble those in finite-dimensional
dynamical systems. In the end, we also present a numerical example
which confirms our analytical predictions.

\section{Stability results for transcritical bifurcations of solitary waves}
We consider the generalized nonlinear Schr\"odinger (GNLS) equations with
arbitrary forms of nonlinearity and external potentials in arbitrary spatial dimensions.
These equations can be written as
\begin{equation}  \label{e:U}
iU_t+\nabla^2 U+F(|U|^2, \x)U=0,
\end{equation}
where $\nabla^2=\partial^2/\partial x_1^2+\partial^2/\partial
x_2^2+\cdots + \partial^2/\partial x_N^2$ is the Laplacian in the
$N$-dimensional space $\textbf{x}=(x_1, x_2, \cdots, x_N)$, and
$F(\cdot, \cdot)$ is a general real-valued function which includes
nonlinearity as well as external potentials. These GNLS equations
include the Gross-Pitaevskii equations in Bose-Einstein condensates
\cite{BEC} and nonlinear light-transmission equations in linear
potentials and nonlinear lattices
\cite{Kivshar_book,Yang_SIAM,Malomed_nonlinear_lattice,Guenbo_nonlinear_lattice} as special
cases. Notice that these equations are conservative and Hamiltonian.

For a large class of nonlinearities and potentials, this equation
admits stationary solitary waves
\begin{equation}  \label{e:Usoliton}
U(\x,t)=e^{i\mu t}u(\x),
\end{equation}
where $u(\x)$ is a real and localized function in the
square-integrable functional space which satisfies the equation
\begin{equation}  \label{e:u}
\nabla^2u-\mu u+F(u^2, \x)  \hspace{0.05cm} u=0,
\end{equation}
and $\mu$ is a real-valued propagation constant. In these solitary
waves, $\mu$ is a free parameter, and $u(\x)$ depends continuously
on $\mu$. Under certain conditions, these solitary waves undergo
bifurcations at special values of $\mu$. Three major types of
bifurcations, namely, saddle-node, pitchfork and transcritical
bifurcations, have been classified in \cite{Yang_classification}. In
addition, linear stability of solitary waves near saddle-node and
pitchfork bifurcations has been determined in
\cite{Yang_saddle1,Yang_saddle2,Weinstein_2008,
Sacchetti_2009,Kirr_2011,Yang_pitchfork}. In this paper, we
determine the linear stability of solitary waves near transcritical
bifurcations in the GNLS equations (\ref{e:U}).

A transcritical bifurcation in the GNLS equations (\ref{e:U}) is
where there are two smooth branches of solitary waves $u^\pm(\x;
\mu)$ which exist on both sides of the bifurcation point
$\mu=\mu_0$, and these two branches cross each other at $\mu=\mu_0$.
A schematic solution-bifurcation diagram of transcritical
bifurcations is shown in Fig.~1(a). Analytical conditions for
transcritical bifurcations in Eq. (\ref{e:U}) were derived in
\cite{Yang_classification}. To present these conditions, we
introduce the linearization operator of Eq. (\ref{e:u}),
\begin{equation}  \label{e:L1}
L_1=\nabla^2-\mu+\partial_u[F(u^2, \x)u],
\end{equation}
which is a linear Schr\"odinger operator. We also
introduce the standard inner product of functions,
\begin{equation*}
\langle f, g\rangle  = \int_{-\infty}^\infty f^*(\x) \hspace{0.05cm}
g(\x) \hspace{0.07cm} d \x,
\end{equation*}
where the superscript `*' represents complex conjugation. In
addition, we define the power of a solitary wave $u(\x;\mu)$ as
\begin{equation*}
P(\mu)=\langle u, u\rangle = \int_{-\infty}^\infty u^2(\x; \mu)
\hspace{0.07cm} d \x,
\end{equation*}
and denote the power functions of the two solitary-wave branches as
\begin{equation*}
P_\pm(\mu)\equiv \langle u^\pm(\x; \mu), u^\pm(\x; \mu) \rangle.
\end{equation*}

If a bifurcation occurs at $\mu=\mu_0$, by denoting the
corresponding solitary wave and the $L_1$ operator as
\begin{equation*}
u_0(\x) \equiv u(\x; \mu_0), \quad L_{10}  \equiv  L_1|_{\mu=\mu_0,\
u=u_0},
\end{equation*}
then $L_{10}$ should have a discrete zero eigenvalue. This is a
necessary condition for all bifurcations. In
\cite{Yang_classification}, the following sufficient conditions for
transcritical bifurcations were derived. In this derivation, it was
assumed implicitly that the function $F(u^2, \x)$ is infinitely
differentiable with respect to $u$.

\vspace{0.3cm} \textbf{Theorem 1} \ Assume that zero is a simple
discrete eigenvalue of $L_{10}$ at $\mu=\mu_0$. Denoting the real
localized eigenfunction of this zero eigenvalue as $\psi(\x)$, and
denoting
\begin{equation} \label{n:G}
G(u;\x) \equiv F(u^2;\x)u, \quad G_2(\x)\equiv \partial_u^2 G|_{u=u_0},
\end{equation}
then if
\begin{equation*}
\langle u_0, \psi\rangle =0, \quad \langle G_2,
\psi^3\rangle \ne 0,
\end{equation*}
and
\begin{equation*}
\Delta \equiv \langle 1-G_2L_{10}^{-1}u_0,,
\psi^2\rangle^2 - \langle G_2, \psi^3\rangle \langle G_2
(L_{10}^{-1}u_0)^2-2L_{10}^{-1}u_0, \psi\rangle > 0,
\end{equation*}
a transcritical bifurcation occurs at $\mu=\mu_0$.

Perturbation series for the two solution branches $u^\pm(\x; \mu)$
near a transcritical bifurcation point were also derived in
\cite{Yang_classification}. It was found that
\begin{equation}
u^\pm(\x; \mu)=u_0(\x)+(\mu-\mu_0)u_1^\pm(\x)+O\{(\mu-\mu_0)^2\},
\end{equation}
where
\begin{equation} \label{e:u1pm}
u_1^\pm=L_{10}^{-1}u_0+b_1^\pm \psi,
\end{equation}
and
\begin{equation}  \label{s:c1case3}
b_1^\pm \equiv \frac{\langle 1-G_2L_{10}^{-1}u_0,
\psi^2\rangle\pm \sqrt{\Delta}}{\langle G_2, \psi^3\rangle }.
\end{equation}
From these perturbation series solutions, power functions $P_\pm(\mu)$ near the bifurcation point can be calculated. In particular, one finds
that
\begin{equation*}
P_+'(\mu_0)=P_-'(\mu_0),
\end{equation*}
thus power curves $P_\pm(\mu)$ of the two solution branches
$u^\pm(\x; \mu)$ have the same slope at the bifurcation point.
Because of this, the two power curves $P_\pm(\mu)$ are tangentially
touched at a transcritical bifurcation. This feature of the
power-bifurcation diagram is shown schematically in Fig.~1(b).
Notice that this power-bifurcation diagram of the transcritical
bifurcation looks quite different from the solution-bifurcation
diagram in Fig.~1(a).

\begin{figure}[h!]
\centerline{\includegraphics[width=0.45\textwidth]{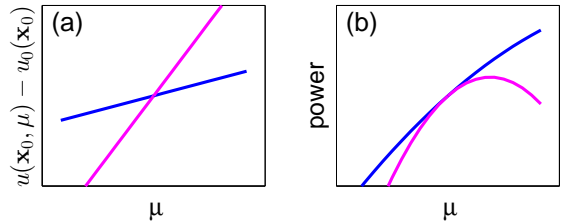}}
\caption{  Schematic diagrams of a transcritical bifurcation: (a)
solution-bifurcation diagram (plotted is the deviation function $u(\x_0;
\mu)-u_0(\x_0)$ versus $\mu$ at a representative $\x_0$ position); (b) power-bifurcation diagram.
The same color represents the same solution branch.}
\end{figure}

The goal of this paper is to analytically determine the linear
stability of solitary waves near a transcritical bifurcation point.
To study this linear stability, we perturb the solitary waves by
normal modes and obtain the following eigenvalue problem (see
\cite{Yang_SIAM}, p176)
\begin{equation} \label{e:LPhi}
{\cal L}\Phi=\lambda \Phi,
\end{equation}
where
\begin{equation}  \label{def:calL}
{\cal L} =  i\left[\begin{array}{cc} 0 & L_0 \\ L_1 & 0
\end{array}\right], \quad \Phi = \left[\begin{array}{c} v \\ w
\end{array}\right],
\end{equation}
\begin{equation} \label{e:L0}
 L_0= \nabla^2-\mu+F(u^2, \x),
\end{equation}
and $L_1$ has been defined in Eq. (\ref{e:L1}). In the later text,
operator ${\cal L}$ will be called the linear-stability operator. If
this linear-stability eigenvalue problem admits eigenvalues
$\lambda$ whose real parts are positive, then the corresponding
normal-mode perturbation exponentially grows, hence the solitary
wave $u(\x)$ is linearly unstable. Otherwise it is linearly stable.
Notice that eigenvalues of this linear-stability problem always
appear in quadruples $(\lambda, -\lambda, \lambda^*, -\lambda^*)$
when $\lambda$ is complex, or in pairs $(\lambda, -\lambda)$ when
$\lambda$ is real or purely imaginary.

Using the operator $L_0$, the solitary wave equation (\ref{e:u}) can
be written as
\begin{equation}  \label{e:L0u}
L_0u=0.
\end{equation}
In particular, when we denote $L_0$ at the bifurcation point as
\begin{equation*}
L_{00} \equiv L_0|_{\mu=\mu_0,\ u=u_0},
\end{equation*}
then
\begin{equation}  \label{e:L00u0}
L_{00}u_0=0,
\end{equation}
thus zero is a discrete eigenvalue of $L_{00}$.

The main result of this paper is the following theorem on
linear-stability eigenvalues of solitary waves near a transcritical
bifurcation point.

\vspace{0.3cm} \textbf{Theorem 2} \
At a transcritical bifurcation point $\mu=\mu_0$ in the GNLS equation (\ref{e:U}), suppose zero is a simple
discrete eigenvalue of operators $L_{00}$ and $L_{10}$,  and
\begin{equation} \label{c:transcritical}
\langle \psi, L_{00}^{-1}\psi\rangle \ne 0, \quad P_\pm'(\mu_0)\ne 0,
\end{equation}
where $\psi$ is the real discrete eigenfunction of the zero eigenvalue in $L_{10}$ (see Theorem 1),
then a single pair of non-zero
eigenvalues $\pm \lambda$ in the linear-stability operator ${\cal L}$ bifurcate out along the real
or imaginary axis from the origin when $\mu\ne \mu_0$. In addition, the
bifurcated eigenvalues $\lambda^\pm$ on the two solution branches $u^\pm(\x;
\mu)$ are given asymptotically by
\begin{equation}  \label{f:lambdapm2case3a}
(\lambda^\pm)^2 \to \mp \gamma (\mu-\mu_0), \quad \mu \to \mu_0,
\end{equation}
where the real constant $\gamma$ is
\begin{equation}
\gamma = \frac{\sqrt{\Delta}}{\langle \psi,
L_{00}^{-1}\psi\rangle}.
\end{equation}

\vspace{0.2cm} A direct consequence of Theorem 2 is the following
Theorem~3 which summarizes the qualitative linear-stability
properties of solitary waves near a transcritical bifurcation point.

\vspace{0.1cm} \textbf{Theorem 3} \ Suppose at a transcritical
bifurcation point $\mu=\mu_0$, the solitary wave $u_0(\x)$ is
linearly stable; and when $\mu$ moves away from $\mu_0$, no complex
eigenvalues bifurcate out from non-zero points on the imaginary
axis. Then under the same assumptions of Theorem 2, both solution
branches $u^\pm(\x; \mu)$ undergo stability switching at the
transcritical bifurcation point. In addition, the two solution
branches have opposite linear stability.

\vspace{0.1cm} Based on this theorem, schematic stability diagrams
for a transcritical bifurcation are displayed in Fig.~2. The
stability behavior in Fig. 2(a) (for solution bifurcation) closely
resembles that for transcritical bifurcations of fixed points in
finite-dimensional dynamical systems \cite{GH}. But the
power-bifurcation diagram (with stability information) in Fig. 2(b)
has no counterpart in finite-dimensional dynamical systems.

\begin{figure}[h!]
\centerline{\includegraphics[width=0.45\textwidth]{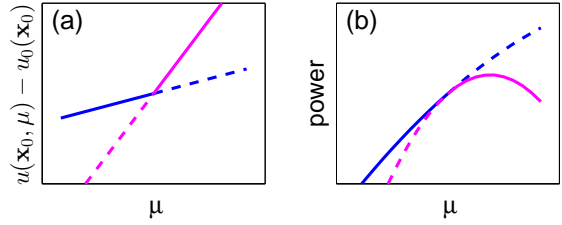}}
\caption{  Schematic diagrams of a transcritical bifurcation with stability information indicated: (a)
solution-stability diagram; (b) power-stability diagram. The same color represents the same solution branch, and
solid (dashed) lines are stable (unstable). }
\end{figure}

Note that for positive solitary waves in the GNLS equations
(\ref{e:U}), linear-stability eigenvalues are all real or purely
imaginary (see \cite{Yang_SIAM}, Theorem 5.2, p176). In addition,
zero is always a simple eigenvalue of $L_{00}$ \cite{Struwe}. In
this case, if zero is also a simple eigenvalue of $L_{10}$ and the
solitary wave $u_0(\x)$ at the bifurcation point is linearly stable,
then under the generic conditions (\ref{c:transcritical}), Theorem 3
applies, thus both solution branches $u^\pm(\x; \mu)$ undergo
stability switching at a transcritical bifurcation point, and  the
two solution branches have opposite linear stability. Such an
example will be given in Sec. \ref{sec:num}.

\section{Proofs of stability results}

\textbf{Proof of Theorem 2} \
First we see from Eqs. (\ref{def:calL}) and (\ref{e:L0u}) that zero is a discrete eigenvalue of $\cal L$ for all $\mu$ values.
At the bifurcation point $\mu=\mu_0$, we further have $L_{10}\psi=0$, thus
\begin{equation} \label{e:L0zeromode}
{\cal L}_0 \left[\begin{array}{c} 0 \\ u_0 \end{array}\right]={\cal
L}_0 \left[\begin{array}{c} \psi \\ 0 \end{array}\right]=0.
\end{equation}
Following the same analysis as in \cite{Yang_pitchfork}, we can
readily show that the algebraic multiplicity of the zero eigenvalue
in $\cal L$ is four at $\mu=\mu_0$ and drops to two when
$0<|\mu-\mu_0|\ll 1$, thus a pair of eigenvalues bifurcate out from
the origin when $\mu\ne \mu_0$. This pair of eigenvalues must
bifurcate along the real or imaginary axis since eigenvalues of
$\cal L$ would appear as quadruples if this bifurcation were not
along these two axes. Next we calculate this pair of eigenvalues
near the bifurcation point $\mu=\mu_0$ by perturbation methods.

The bifurcated eigenmodes on the solution branches $u^\pm(\x;
\mu)$ have the following perturbation series expansions,
\begin{eqnarray}
&& v^\pm(\x;\mu)=\sum_{k=0}^\infty (\mu-\mu_0)^{k} v_k^\pm (\x),  \\
&& w^\pm(\x; \mu)  = \lambda_0^\pm (\mu-\mu_0)^{1/2}\sum_{k=0}^\infty
(\mu-\mu_0)^{k}
w_k^\pm(\x),   \label{e:wexpand} \\
&& \lambda^\pm(\mu) =
i\lambda_0^\pm (\mu-\mu_0)^{1/2}\left(1+\sum_{k=1}^\infty (\mu-\mu_0)^{k}
\lambda_k^\pm \right).           \label{e:lambdaexpand}
\end{eqnarray}
We also expand $L_0$ and $L_1$ on the solution branches $u^\pm(\x;
\mu)$ into perturbation series
\begin{eqnarray}
&& L_0^\pm=L_{00}+(\mu-\mu_0)L_{01}^\pm+(\mu-\mu_0)^2L_{02}^\pm +\dots, \\
&& L_1^\pm=L_{10}+(\mu-\mu_0)L_{11}^\pm+(\mu-\mu_0)^2L_{12}^\pm +\dots.   \label{e:L1expand}
\end{eqnarray}
Substituting the above perturbation expansions into the
linear-stability eigenvalue problem (\ref{e:LPhi}) and at various orders of $\mu-\mu_0$, we get a sequence of
linear equations for $(v_k, w_k)$:
\begin{eqnarray}
&& \hspace{-0.5cm}
L_{10}v_0^\pm=0,    \label{e:L10v0} \\
&& \hspace{-0.5cm}
L_{00}w_0^\pm=v_0,   \label{e:L00w0}  \\
&& \hspace{-0.5cm}
L_{10}v_1^\pm=\left(\lambda_0^\pm\right)^2w_0-L_{11}^\pm v_0,   \label{e:L10v1}
\end{eqnarray}
and so on.

First we consider the $v_0^\pm$ equation (\ref{e:L10v0}). In view of the
assumption in Theorem 2, the only solution to this equation (after
eigenfunction scaling) is
\begin{equation}
v_0^\pm=\psi.
\end{equation}
For the inhomogeneous $w_0^\pm$ equation (\ref{e:L00w0}), it admits
a single homogeneous solution $u_0$ due to (\ref{e:L00u0}) and the
assumption in Theorem 2. Since $L_{00}$ is self-adjoint and $\langle
u_0, \psi\rangle=0$ for transcritical bifurcations (see Theorem 1),
the Fredholm condition for Eq. (\ref{e:L00w0}) is satisfied, thus
this equation admits a real localized particular solution
$L_{00}^{-1}\psi$, and its general solution is
\begin{equation} \label{f:w0}
w_0^\pm=L_{00}^{-1}\psi + c_0^\pm u_0,
\end{equation}
where $c_0^\pm$ is a constant to be determined from the solvability
condition of the $w_1^\pm$ equation later.

For the $v_1^\pm$ equation (\ref{e:L10v1}), it is solvable if and only
if its right hand side is orthogonal to the homogeneous solution $\psi$. Utilizing the $v_0^\pm$
and $w_0^\pm$ solutions derived above and recalling $\langle u_0, \psi\rangle=0$, this orthogonality condition
yields the formula for the eigenvalue coefficient $\lambda_0^\pm$ as
\begin{equation}  \label{f:lambda0case3}
(\lambda_0^{\pm})^2=\frac{\langle \psi, L_{11}^\pm
\psi\rangle}{\langle \psi, L_{00}^{-1}\psi\rangle}.
\end{equation}
Due to notations (\ref{n:G}) and the definition (\ref{e:L1}) for
$L_1$, it is easy to see that $L_{11}^\pm$ in the expansion
(\ref{e:L1expand}) is
\begin{equation}
L_{11}^\pm=G_2u_1^\pm-1,
\end{equation}
where $u_1^\pm$ is given in Eq. (\ref{e:u1pm}). Inserting this
$L_{11}^\pm$ into (\ref{f:lambda0case3}), we find that
\begin{equation}
(\lambda_0^{\pm})^2 = \pm \frac{\sqrt{\Delta}}{\langle \psi,
L_{00}^{-1}\psi\rangle}.
\end{equation}
Substituting this formula into (\ref{e:lambdaexpand}), we then
obtain the asymptotic expression for the eigenvalues
$(\lambda^\pm)^2$ as (\ref{f:lambdapm2case3a}) in Theorem 2. This completes the proof of Theorem 2. $\Box$

\textbf{Proof of Theorem 3} \ Under conditions of Theorem 3, when
$\mu$ moves away from $\mu_0$, the only instability-inducing
eigenvalue bifurcation is from the origin. We have shown in Theorem
2 that this zero-eigenvalue bifurcation creates a single pair of
eigenvalues whose asymptotic expressions are given by Eq.
(\ref{f:lambdapm2case3a}). This formula shows that on the same
solution branch (i.e., $u^+(\x; \mu)$ or $u^-(\x; \mu)$), if the
bifurcated eigenvalues are real (unstable) on one side of
$\mu=\mu_0$, then they are purely imaginary (stable) on the other
side of $\mu=\mu_0$. Thus stability switching occurs at the
bifurcation point $\mu=\mu_0$ for both solution branches. This
formula also shows that at the same $\mu$ value, if the bifurcated
eigenvalues are real on one solution branch, then they would be
purely imaginary on  the other solution branch. Thus the two
solution branches always have opposite linear stability. This
completes the proof of Theorem 3. $\Box$

\section{A numerical example}  \label{sec:num}

An example of transcritical bifurcations in the GNLS equation (\ref{e:U}) has been reported in \cite{Yang_classification}. This example is
\begin{equation}  \label{e:Uexample3}
iU_t+U_{xx}-V(x)U+|U|^2U-0.2 |U|^4U +\kappa_c |U|^6U=0,
\end{equation}
where $V(x)$ is an asymmetric double-well potential
\begin{equation} \label{e:Vexample2}
V(x)=-3.5\: \mbox{sech}^2(x+1.5)-3\: \mbox{sech}^2(x-1.5),
\end{equation}
and $\kappa_c\approx 0.01247946$. The potential (\ref{e:Vexample2})
is displayed in Fig. 3(a), and the power diagram of this bifurcation
is shown in Fig.~3(b). We see that two smooth power branches, namely
the upper $c_1$-$c_2$ branch and the lower $d_1$-$d_2$ branch,
tangentially connect at the bifurcation point $(\mu_0, P_0)\approx
(3.278, 14.36)$. This tangential connection agrees with the
analytical result on the power diagram (see Fig.~1(b)). Profiles of
solitary waves at the marked $c_1, c_2, d_1, d_2$ locations on this
power diagram are displayed in Fig.~3(c-d), and their
linear-stability spectra are shown in Fig.~3($c_1,c_2,d_1,d_2$)
respectively. These spectra indicate that the solitary waves at
$c_1$ and $d_2$ of the power diagram are linearly stable, whereas
the other two solitary waves at $c_2$ and $d_1$ of the power diagram
are linearly unstable. Thus both the upper $c_1$-$c_2$ branch and
the lower $d_1$-$d_2$ branch switch instability at the bifurcation
point, and the $c_1$-$c_2$ and $d_1$-$d_2$ branches have opposite
linear stability. These numerical results confirm the analytical
results in Theorem 3 (see also Fig.~2(b)).

\begin{figure}[h!]
\centerline{\includegraphics[width=0.9\textwidth]{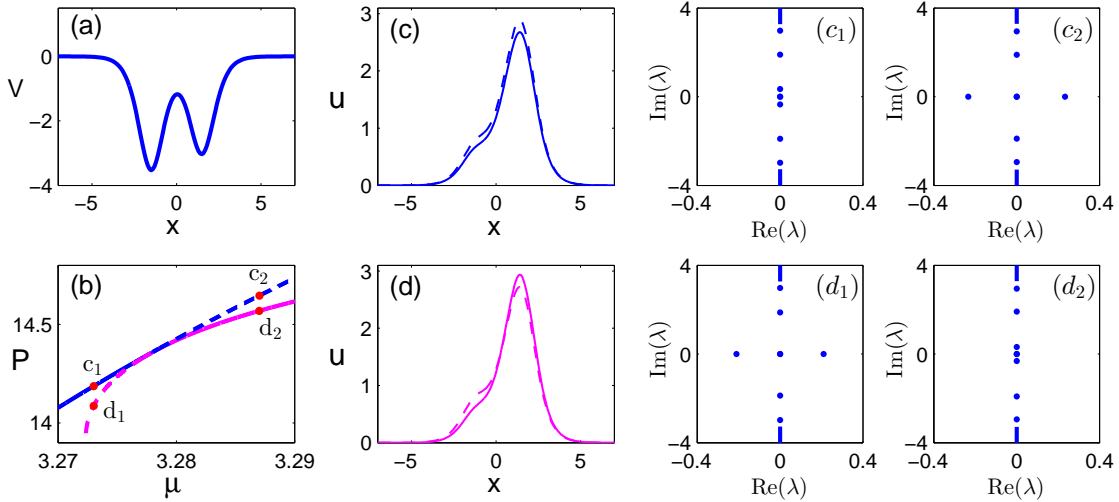}}
\caption{ Stability switching at a transcritical bifurcation in Example (\ref{e:Uexample3}). (a) The asymmetric double-well potential
(\ref{e:Vexample2}); (b) the power diagram; (c) profiles of
solitary waves at locations $c_1$ (solid) and $c_2$ (dashed) of the upper power curve in (b); (d) profiles of solitary waves
at locations $d_1$ (dashed) and $d_2$ (solid) of the lower
power curve in (b). ($c_1,c_2,d_1,d_2$) stability spectra for solitary waves at locations of the same letters on the power diagram (b). }
\end{figure}

\section{Summary and discussion}
In summary, linear stability of solitary waves near transcritical
bifurcations was analyzed for the GNLS equations (\ref{e:U}) with
arbitrary forms of nonlinearity and external potentials in arbitrary
spatial dimensions. It was shown that both solution branches undergo
stability switching at the transcritical bifurcation point. In
addition, the two solution branches have opposite stability.
Analytical formulae for the unstable eigenvalues were also derived.
These analytical stability results were compared with a numerical
example and good agreement was obtained.

The above stability properties closely resemble those in
finite-dimensional dynamical systems,  where it is well known that
the stability of fixed-point branches near a transcritical
bifurcation point exhibits the same behaviors as above \cite{GH}.
However, this happy resemblance, which we proved in this paper,
should not be taken for granted. Indeed, it has been shown that for
saddle-node and pitchfork bifurcations, stability properties in the
GNLS equations differ significantly from those in finite-dimensional
dynamical systems \cite{Yang_saddle1,Yang_saddle2,Yang_pitchfork}.
For instance, at a saddle-node bifurcation point, there is no
stability switching in the GNLS equations (\ref{e:U})
\cite{Yang_saddle1,Yang_saddle2}, but any dynamical-system textbook
would say that such stability switching takes place \cite{GH}. Thus
it may be more appropriate to view this similar stability on
transcritical bifurcations in the GNLS equations and
finite-dimensional dynamical systems as a happy surprise rather than
a trivial expectation.

Another approach to qualitatively study the linear stability of
nonlinear waves in Hamiltonian systems is the Hamiltonian-Krein
index theory
\cite{Kapitula_index,Kapitula_appendix,Peli_index,Peli_index_2010,Kapitula_Promislow}.
In this approach, the number of unstable eigenvalues in the
linear-stability operator ${\cal L}$ is related to the number of
positive eigenvalues in operators $L_0$ and $L_1$ under appropriate
conditions. Near a transcritical bifurcation point $\mu=\mu_0$, we
can show that the zero eigenvalue in $L_{10}$ bifurcates out as
\begin{equation*}
\Lambda_\pm(\mu)\to \frac{\pm \sqrt{\Delta}}{\langle \psi, \psi\rangle} \, (\mu-\mu_0), \qquad \mu\to \mu_0,
\end{equation*}
where $\Lambda_\pm(\mu)$ is the eigenvalue of $L_1$ on the
$u^\pm(\x; \mu)$ solution branch. Using this formula, the
qualitative stability results in Theorem 3 can be reproduced by the
index theory (as was done in \cite{Yang_pitchfork} for pitchfork
bifurcations). However, this index-theory approach requires more
restrictive conditions on the spectra of $L_0$ and $L_1$ operators
\cite{Kapitula_index,Peli_index_2010,Yang_pitchfork}, and it cannot
yield quantitative linear-stability eigenvalue formula
(\ref{f:lambdapm2case3a}) either.

It should be recognized that transcritical bifurcations in the GNLS
equations (\ref{e:U}) occur less frequently than saddle-node or
pitchfork bifurcations. Indeed, in the example (\ref{e:Uexample3}),
if the seventh-power coefficient $\kappa$ is not equal to that
special value $\kappa_c$, this transcritical bifurcation would
either turn into a pair of saddle-node bifurcations or disappear,
depending on whether $\kappa$ is less than $\kappa_c$ or greater
than $\kappa_c$. Due to this less frequent occurrence, one might
wonder how useful the stability results in this paper are. To
address this concern, it is helpful to view a transcritical
bifurcation as the limit when two saddle-node bifurcations coalesce
with each other (such as when $\kappa \to \kappa_c^-$ in the example
(\ref{e:Uexample3})). In this connection, the stability results
obtained in this paper for transcritical bifurcations can also be
used to help determine stability properties of nearby saddle-node
solution branches. Thus the stability results in this article can be
useful beyond transcritical bifurcations.

\section*{Acknowledgment}
This work is supported in part by the Air Force Office of Scientific
Research (Grant USAF 9550-12-1-0244) and the National Science
Foundation (Grant DMS-0908167).

\bigskip

\end{document}